\documentclass[letterpaper,aps,prl,twocolumn,showpacs,preprintnumbers,superscriptaddress]{revtex4}
\usepackage{graphicx}
\usepackage{dcolumn}
\usepackage{bm}
\usepackage{amssymb}
\usepackage{amsmath}
\usepackage{times}
\usepackage[usenames,dvipsnames]{color}

\begin{document}

\title{Glassy dynamics, aging and thermally activated avalanches in
interface pinning at finite temperatures}

\author{Jos\'e J. Ramasco}\email{jose.ramasco@emory.edu}\affiliation{Physics
Department, Emory University, Atlanta GA 30322, USA}

\author{Juan M. L\'opez}
\affiliation{Instituto
de F\'{\i}sica de Cantabria (IFCA), CSIC-UC, E-39005 Santander,
Spain}

\author{Miguel A. Rodr\'{\i}guez}
\affiliation{Instituto
de F\'{\i}sica de Cantabria (IFCA), CSIC-UC, E-39005 Santander,
Spain}

\date{\today}

\begin{abstract}
We study numerically the out-of-equilibrium dynamics of interfaces at finite
temperatures when driven well below the zero-temperature depinning threshold.
We go further than previous analysis by including the most relevant
non-equilibrium correction to the elastic Hamiltonian. We find that the
relaxation dynamics towards the steady-state shows glassy behavior, aging and
violation of the fluctuation-dissipation theorem. The interface roughness
exponent $\alpha \approx 0.7$ is found to be robust to temperature changes. We
also study the instantaneous velocity signal in the low temperature regime and
find long-range temporal correlations. We argue $1/f$-noise arises from the
merging of local thermally-activated avalanches of depinning events.
\end{abstract}


\maketitle

The dynamics of interfaces in random media has received much attention, both
theoretical and experimental, in the last decade. This is largely due to the
interest in driven interfaces as models for nonlinear cooperative transport in
many contexts, such as stochastic surface growth and kinetic
roughening \cite{barabasi}, magnetic flux lines in type-II
superconductors \cite{scII}, the motion of charge density
waves \cite{gruner}, propagation of fracture cracks \cite{bouchaud},
fluid imbibition in porous media \cite{alava}, etc.
The competition between elastic interactions and quenched disorder leads to a
rich behavior, which is not yet fully understood. A key macroscopic observable
to look at is the average interface velocity $v$ as a function of the external
driving force $f$. At zero temperature, quenched disorder tends to pin the
interface and this leads to a depinning transition from a pinned phase ($v=0$)
to a moving phase ($v>0$) at a critical value of the external driving force
$f_c$. However, one particularly important question is how the system responds
to this external forcing in the presence of non-negligible thermal
fluctuations.

At finite temperatures, when the interface is driven well below the
zero-temperature critical force $f_c(T=0)$, the interface velocity $v$ is
extremely small. The interface then moves in bursts of activity, which are
thermally activated and, in the steady state, the response of the system to a
small external force exhibits glassy properties known as {\em creeping}
\cite{creep}. Experiments \cite{creep_exp} and theoretical studies
\cite{creep_theo} have confirmed the existence the creep behavior. Theoretical
approaches to this problem using mean-field or functional renormalization group
techniques \cite{cugliandolo96,chauve,balents,schehr,nattermann} are usually
valid in high dimensions and difficult to extend to one dimensional interfaces.
More importantly, these approaches treat the dynamics of an elastic string in a
random potential, {\it i.e.} interaction terms derive from an elastic energy
$\mathcal{H} = (\nu/2) \int dx[ \sqrt{(\nabla h)^2+1} + \Phi_p(x,h)]$ in a
random pinning potential $\Phi_p$ in the small slopes limit $|\nabla h|\ll 1$.
However, the elastic line approach gives a roughness exponent $\alpha \approx
1.2$ at the depinning transition, which leads to difficulties in the
thermodynamic limit since interface fluctuations $\langle [h(x) - \overline
h(x)]^2\rangle \sim L^{2\alpha}$ diverge and, indeed, the small slopes
approximation breaks down \cite{rosso,doussal}.

Recent theoretical arguments have pointed out that a
Kardar-Parisi-Zhang (KPZ) nonlinearity \cite{kpz} may emerge from
anisotropy of the disorder \cite{tang} and/or from higher order
terms in an elastic expansion \cite{rosso,doussal}. Also, recent
developments in imaging technology have allowed to directly
visualize the boundary between the vortex invaded and the vortex
free (Meissner) regions in superconductors. In a series of nice
experiments Surdeanu {\it et al.} \cite{surdeanu} recently measured
the temporal and spatial correlations of these fronts for a
YBa$_2$Cu$_3$O$_{7-x}$ superconductor. They reported interfaces that
exhibited scale-invariant kinetic roughening with exponents
consistent with KPZ dynamics. Similar results were later obtained in
Ref. \cite{vlasko} for thin Nb superconductor films. These
theoretical and experimental results make evident the importance of
understanding the effect of thermal fluctuations on pinned
interfaces when the interaction between degrees of freedom contains
nonequilibrium contributions, like the KPZ nonlinearity.

In this Letter we study the relaxation towards the steady state of an interface
locally pinned by a disordered background. Our study includes the most relevant
KPZ non-equilibrium correction to the elastic energy. We focuss on the
interface dynamics well below the depinning transition, when only temperature
can locally detach the interface. We find that the system very slowly relaxes
(creeping) to a steady-state. During the relaxation regime the system exhibits
aging and violation of the fluctuation-dissipation theorem. The roughness
exponent of the string is found to be $\alpha \approx 0.7$ at short scales and
robust to changes in temperature. Remarkably, the instantaneous velocity signal
exhibits long-range temporal correlations. We argue $1/f$-noise arises from the
merging of local thermally-activated avalanches of depinning events.

We consider a 1D line described by a single-valued function $h(x,t)$, giving
the transversal position $h$ of the front from the $h=0$ axis at time $t$. The
front is moving through a heterogeneous medium, which can be described by a
quenched disorder $\eta(x,h)$ and the interface obeys the equation of motion:
\begin{equation}
\label{tqkpz}
\partial_t h = \nu \, \nabla^2 h + \lambda \, (\nabla h)^2 + \sqrt{T} \, \xi(x,t) +
\sigma \, \eta(x,h) + f,
\end{equation}
where $\xi(x,t)$ is a gaussian white noise with zero mean and unit variance
describing the thermal fluctuations at temperature $T$, $\nu$ is the elastic
constant, and the KPZ nonlinearity $\lambda \, (\nabla h)^2$ gives the most
relevant nonequilibrium correction to the equilibrium elastic energy.  We only
consider here the case of non-correlated gaussian disorder so that $\langle
\eta(x,y) \eta(x',y') = \delta(x-x') \, \delta(y-y')$ and $\langle \eta (x,y)
\rangle = 0$.

\begin{figure}
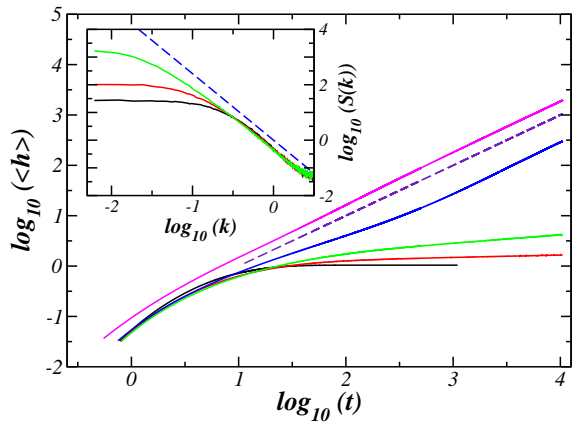

\centerline{\includegraphics *[width=75mm]{fig1.eps}}
\caption{(Color
online) Average height as a function of time for the nonlinear
equation. The different curves correspond from bottom to top to the
temperatures $T = 0, 0.01, 0.04, 0.16$ and $0.64$. In the inset, the
spatial power spectra of the interfaces for the three lowest
temperatures (colors match in both graphs). The dashed lines have
slope $1$ in the main plot and slope $-(2\alpha +1) = - 2.4$ in the
inset.} \label{hm}
\end{figure}

\paragraph {Interface scaling.-} We have carried out extensive numerical
simulations of the Langevin equation of motion (\ref{tqkpz}). To solve
(\ref{tqkpz}) numerically we discretized $h(x,t)$ along the $x$ direction with
lattice spacing $\Delta x = 1$ and used a stochastic Euler scheme to integrate
the equation of motion with periodic boundary conditions. The constants $\nu$
and $\sigma$ can be scaled out and without lose of generality we have set $\nu
= \sigma = 1$. All the results reported in the following were obtained for an
integration time step $\Delta t = 0.01$, which was enough to guaranty stability
of the numerical scheme and $\Delta t$-independence of the numerical results.
We are interested in the fully nonlinear regime, thus, for the system sizes we
used (up to $L=2046$), the strength of the nonlinear term is set to $\lambda =
1$. Our results do not depend on $\lambda$, as far as it is large enough to
allow the nonlinearity to fully operate well before saturation is reached. For
comparison we have also studied the usual elastic regime for $\lambda = 0$.

Simulations at $T=0$ revealed that the depining transition takes
place at $f_c(T=0,\lambda = 1) = 0.36 \pm 0.01$ and $f_c(T=0,\lambda
= 0) = 0.99 \pm 0.01$ for the nonlinear and linear models,
respectively. A more precise determination of the critical
thresholds is not required since we are interested in the dynamics
of the system well below the critical point. In the following,
unless otherwise stated, we will always be driving the system with
external forces $f(\lambda = 1) = 0$ and $f(\lambda = 0) = 0.1$,
well below the critical threshold. We now focuss on the effects of
varying the external temperature $T$ when the system is in the
zero-temperature pinned phase.

We always start our simulation with a flat interface $h=0$ and then let the
system evolve at a given temperature $T$. We monitor the average height
$\langle \overline{h} \rangle$ from which the interface velocity can be
obtained in the long times limit
$v \equiv \lim_{t \to \infty} \langle \overline{h} \rangle/t$.

Initially, as may be seen in Figure 1, the front advances very fast. The
interface moves through the quenched disorder and progressively gets locally
trapped at more and more pinning sites. The duration of this transient is
system size independent and relatively short. After that initial outburst, the
interface begins relaxation towards the steady state. The latter is
characterized by a constant velocity. The time span of the relaxation process
strongly depends on temperature, increasing very dramatically at low $T$.
During relaxation, the average height displays a nonlinear growth in time,
$\langle \overline{h} \rangle \sim t^{\gamma(T)}$, characteristic of the
creeping regime. The actual value of $\gamma(T)$ decays rapidly with
decreasing $T$, being the average height at very low $T$ compatible with a
logarithmic growth.

This singular dynamics of the average height during the creeping
regime imposes constrains upon the evolution of the interface width.
Since $w(t,L) \sim t^{\alpha/z}$ cannot grow faster than the height
$\langle \overline{h} \rangle \sim t^\gamma$, we have $\alpha/z \leq
\gamma(T)$. Simulations show that the value of the roughness
exponent $\alpha \approx 0.7$ remains essentially constant at low
temperatures. In the inset of Figure 1 we plot the structure factor
$S(k,t) = \langle \hat{h}(k,t) \hat{h}(-k,t)\rangle$ at long times
$t=10^4$ for three different temperatures $T=0, 0.01$ and $0.04$
(where $\hat{h}$ is the the Fourier transform of the interface
height in a system of lateral size L, with $k$ being the spatial
frequency in reciprocal space). The inset clearly shows that ({\it
i}) after long times $10^4$ the interface has not yet reached a
steady state, ({\it ii}) the interface is scale-invariant $S(k,t)
\sim k^{-(2\alpha+1)}$ at short scales $k\gg t^{-1/z} $, and ({\it
iii}) the roughness exponent $\alpha$ does not depend on
temperature. This in turn implies that it is the dynamic exponent
$z$ the one changing with temperature (since $\alpha/z \leq
\gamma(T)$). This is made more evident as $T \to 0$, when dynamics
is dramatically slowed down and $z$ suffers a significative
increment.

\paragraph{Aging.-} Ultra-slow relaxation towards the steady state indicates typical
out of equilibrium glassy features in the system, in particular the existence
of aging--namely, the breakdown of the time translation invariance symmetry
\cite{cugliandolo96,yoshino,cugliandolo97}. In order to study aging in our system we
compute the two-times connected correlation function
\begin{equation}
\Gamma(t,t_w) = \langle \overline{[h(x,t)-\langle h \rangle (t)] \, [h(x,t_w) -
\langle h \rangle (t_w)]} \rangle.
\end{equation}
If the system ages, $\Gamma(t,t_w)$ must be a function of both $t$
and $t_w$, and not only of their difference $t-t_w$. We follow the
standard procedure, which closely mimics what can be done in real
experiments. The interface is evolved at some very high fixed
temperature (we typically used $T = 4$ for the nonlinear equation
and $T = 36$ for the linear case) to anneal the system until the
stationary state is reached. Then, when the system is in the
steady-state, the temperature is suddenly decreased to the desired
measuring level $T$. This sets up the system out of equilibrium and
the time counter is started, $t=0$, at the freezing event. We then
monitor the relaxation dynamics by measuring $\Gamma(t,t_w)$.
Typical $\Gamma(t,t_w)$ curves for a set of waiting times $t_w$ are
shown in Figure 2 for a external temperature $T=0.04$ and $0.49$ for
the nonlinear and linear cases, respectively. Both the nonlinear and
the linear equations show clear aging during relaxation towards the
steady state similarly to other glassy systems.

\begin{figure}
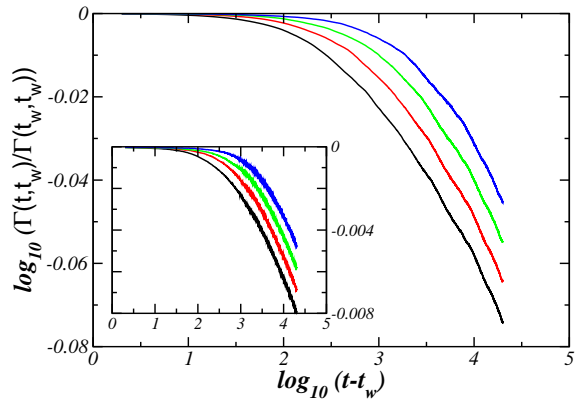

\centerline{\includegraphics *[width=75mm]{fig2.eps}}
\caption{(Color online) Connected correlation as a function of the
time difference $t-t_w$. The main graph is for the nonlinear
equation at $T = 0.04$ and $t_w = 250, 500, 1000$ and $2000$ from
left to right. In the inset, the same plot but for the linear
equation, at $T = 0.49$ and the same set of $t_w$ values.}
\label{Gt}
\end{figure}

\paragraph{Fluctuation-dissipation violation.-} Another important aspect of glassy
dynamics is the way in which the fluctuation-dissipation relation is broken. At
equilibrium, the susceptibility is linearly related to the spatial correlations
of the conjugated variable. A modification for slowly evolving out of
equilibrium systems has been recently proposed by Cugliandolo, Kurchan and
Peliti \cite{cugliandolo97}, through the generalized response function
\begin{equation}
\label{fdt}
R(t,t_w) = \frac{X[\Gamma]}{T} \, \frac{\partial \Gamma (t,t_w)}{\partial t_w} ,
\end{equation}
where $R(t,t_w) = \langle \overline{ \delta h(x,t)/\delta f(x,t_w)} \rangle$,
i.e., the response of the conjugated variable (the interface height in our
case) at a time $t$ to the application of a local external field (the driving
force $f$ in our case) at the same position $x$ at a previous time $t_w$.
$X[\Gamma]$ is a functional that can vary between two asymptotic constant
values, either it takes value one (for scales that are already in the
stationary regime) or $X[\Gamma] \to X_\infty$ (for those scales that are still
out of equilibrium).

\begin{figure}
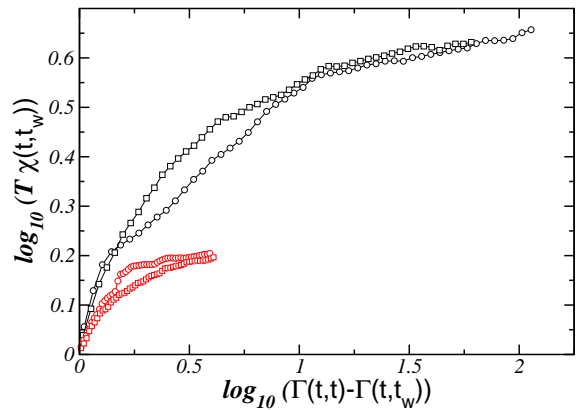

\centerline{\includegraphics *[width=75mm]{fig3.eps}}
\caption{(Color online) Parametric representation of the
susceptibility as a function of the connected correlations for the
nonlinear model with $T = 0.01$ (black curves) and $0.02$ (red
ones). The waiting times in both cases are $t_w = 500$ (circles) and
$t_w = 1000$ (squares).} \label{fdt-fig}
\end{figure}

Numerically, it is more convenient to study the susceptibility, defined as
$\chi(t,t_w) = \int_{t_w}^t ds \, R(t,s) $, instead of the response function
$R(t,t_w)$. The reason being that the computation of $\chi(t,t_w)$ does not
require the introduction of an instantaneous spike-like perturbation
in the external field and, in addition, the integration in time implies
that the external field operates during a period of time making the estimation
of its effects much easier to detect. We integrate $R(t,s)$ in Eq.\ (\ref{fdt})
in the variable $s$ and obtain
\begin{equation} \chi(t,t_w) =
\frac{X[\Gamma]}{T} \, [\Gamma(t,t) - \Gamma(t,t_w)] .
\end{equation}
In order to check whether the fluctuation-dissipation relation is
broken during relaxation in our system, we use a similar numerical
setup as before. We anneal the system by heating at very high
temperatures until a steady-state is reached, then the temperature
is decreased and that event marks the time origin $t=0$. After a
waiting time $t_w$, we run in parallel two copies A and B of the
interface. Copy A is then driven with constant force $f^{\rm A}(x,t)
= f^{\rm A}$ as before, and copy B is driven with a slightly
different local driving force $f^{\rm B}(x,t) = f^{\rm A} + \delta
f(x)$, where $\delta f(x)$ is a small uncorrelated random field
which takes at random the values $\pm \epsilon$ with equal
probability. The susceptibility can then be approximated by
$\chi(t,t_w) \sim \langle \overline{(h^{\rm B}(x,t) - h^{\rm
A}(x,t)) \, \delta f(x) } \rangle/|\epsilon|$. In Figure 3,
parameteric plots with the susceptibility as a function of
$\Gamma(t,t)-\Gamma(t,t_w)$ are shown for the nonlinear equation
($\lambda = 1$) and for different temperatures. The
fluctuation-dissipation theorem is clearly violated, again akin to
observations in other glassy systems.

\paragraph{Thermally activated avalanches.-} Finally, we pay attention to the
dynamics of avalanches at finite temperatures in the nonlinear model
Eq.\ (\ref{tqkpz}). Avalanches in forced superconductors have
attracted much attention \cite{altshuler,aval}, but little is known
about avalanche dynamics at a finite $T$. $\overline{h}(t)$ data
(not shown) for a single disorder realization have a dented and
devil staircase-like appearance caused by the stick--slip motion
associated with avalanches triggered by a local thermally-activated
depinning event. We monitor the avalanche dynamics in our system by
looking at the temporal correlations of the average velocity signal
$v(t) = \overline{\partial_t h} (t)$ well below $f_c(T=0)$. In
Figure 4, we plot the spectral density $\mathcal{S}(w) = \langle
|\hat{v}(w)|^2 \rangle$, averaged over disorder realizations. The
velocity signal $v(t)$ shows long-range temporal correlations. In
the low-frequency limit we observe a power-law decay $\mathcal{S}(w)
\sim w^{-\theta}$ with a temperature-dependent exponent $\theta(T)$
ranging from uncorrelated noise $\theta \sim 0$ (at very low
temperatures) to long-range correlations $1.0<\theta(T)<1.2$ for
temperatures in the range $0.05<T<0.20$. For even higher
temperatures, $T \geq 0.25$, velocity spectra clearly show a
crossover to a different power-law decay with exponent $\theta(T)
\approx 0.30$, which then remains unaltered when temperature is
further increased. The latter regime corresponds to a free-moving
interface in the KPZ universality class, which velocity-fluctuation
spectra is known to diverge as $1/w^\theta$ at low frequencies,
where $\theta = (d+4)/z-3$ in $d+1$ dimensions \cite{krug}. In $d=1$
one then finds that $\theta = 1/3$ in good agreement with our
numerical results at high temperatures where the interface is freely
moving with a finite displacement velocity. We obtained similar
results for the linear case ($\lambda=0$), where $\theta \sim 1$ in
the intermediate temperatures regime, and the crossover to $w^0$ for
the highest temperatures, as expected for the linear equation
\cite{krug}.

\begin{figure}
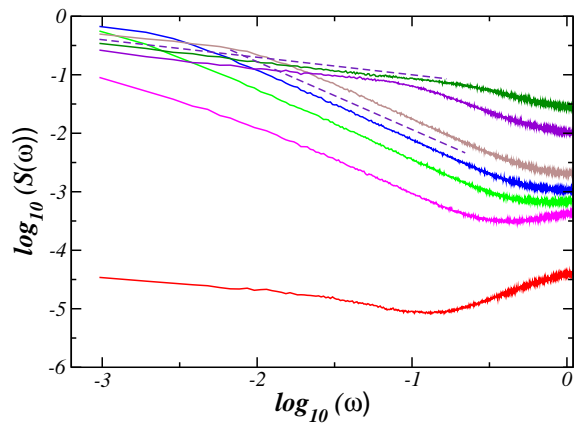

\centerline{\includegraphics *[width=75mm]{fig4.eps}}
\caption{(Color online) Spectral densities of the velocity signal
for the nonlinear equation ($\lambda=1$). Temperatures are $T =
0.01$, $0.09$, $0.1225$, $0.16$, $0.25$, $1$ and $2.25$ from bottom
to top. The straight lines are guide for the eye and have a slope
$-0.3$ and $-1.16$ respectively.}
\end{figure}

This change in the temporal correlations of the velocity signal at
high enough temperatures can be pinned down to the existence of
thermally activated avalanches of depinning events. When the
interface is driven well below $f_c$ at finite temperatures,
uncorrelated avalanches of typical size $\xi$ are triggered all over
the front. At very low temperatures thermally activated avalanches
are very unlikely and localized events, disjoint and uncorrelated to
one another, giving rise to a typical flat spectrum ($\theta \sim
0$) of the velocity signal. However, at higher temperatures we
expect more sites to detach from disorder and so more local
avalanches to occur. Avalanches can overlap and merge to give rise
to a larger avalanche, eventually covering a macroscopic region of
the interface. This leads to bursts of spatially connected moving
events and, consequently, long-range temporal correlations ($\theta
\sim 1$) in the interface velocity fluctuations. This can be seen in
Figure 4 for $0.09 \leq T \leq 0.16$. The higher the temperature,
the more likely a site gets depinned, triggering a local avalanche
of depinning events of size $\xi$. Eventually, at high enough
temperatures ($T \approx 0.25$ in Figure 4) all the pinning sites
are overcome by the merging of local avalanches that can extend
across the whole system freeing the interface from the disorder.
Further increasing the temperature above that point will not affect
the dynamics since temperature is high enough to overcome local
pinning and the depinning transition is actually wiped out. In this
regime the interface is described by the KPZ dynamics and $w^{-1/3}$
velocity-fluctuation spectra.

\begin{acknowledgements}
This work was partially supported by the NSF under grant No. 0312510 (USA) and
by the Ministerio de Educaci\'on y Ciencia (Spain) under projects
BFM2003-07749-C05-03 and CGL2004-02652/CLI.
\end{acknowledgements}

\end{document}